\documentclass[usenatbib,useAMS]{mn2e}
\usepackage{graphicx}
\usepackage{epstopdf}
\usepackage{ctable}
\usepackage{url}
\usepackage{times}

\newcommand{\msun}{~\mathrm{M_{\odot}}}
\newcommand{\msunperyr}{~\mathrm{M_{\odot} {\rm ~yr}^{-1}}}
\newcommand{\pergyr}{~\mathrm{Gyr}^{-1}}

\newcommand{\mstar}{M_{\star}}
\newcommand{\mdust}{M_\mathrm{d}}

\newcommand{\acknowledgments}{\begin{small}\section*{Acknowledgments}\end{small}}

\title[SFR and stellar mass limits for SMGs]{The star formation rate and stellar mass limits for submillimetre galaxies implied by recent interferometric observations}

\author[C.~C. Hayward]{
	\parbox[t]{\textwidth}{
		Christopher C. Hayward$^{1}$\thanks{E-mail: christopher.hayward@h-its.org}}
	\vspace*{6pt} \\
	$^1$Heidelberger Institut f\"ur Theoretische Studien, Schloss-Wolfsbrunnenweg 35, 69118 Heidelberg, Germany \\
}

\usepackage{hyperref}

\begin{document}

\date{Accepted 2013 April 2. Received 2013 March 26; in original form 2013 January 29}

\pagerange{\pageref{firstpage}--\pageref{lastpage}} \pubyear{2013}

\maketitle

\label{firstpage}

\begin{abstract}
Explaining the observed number counts of submillimetre (submm) galaxies (SMGs) has been a longstanding challenge for theoretical
models. Surprisingly, recent observations have suggested that the brightest SMGs
are almost exclusively multiple fainter sources blended into a single source in the single-dish surveys.
This result is in contrast with the predictions of our previously presented theoretical model, which includes some effects of blending.
In this Letter, we consider the implications of an upper limit on the submm flux density for the demographics of the SMG population.
Using a relation amongst submm flux, star formation rate (SFR), and
dust mass ($\mdust$) from our previous work, we infer the maximum SFR for a range of flux density limits. For $\mdust = 2 \times 10^9~(5 \times 10^8)
\msun$, the SFR limit that corresponds to an 870-\micron~flux density ($S_{870}$) limit in the range $9-12.5$ mJy is in the range $\sim630-1400~(3600-7700) \msunperyr$.
The SFR limit implies a correspondingly sharp, redshift-dependent cutoff in the stellar mass ($\mstar$) function, the value of which we predict using
the $S_{870}$--$\mstar$ relation predicted by our model. The $\mstar$ limit decreases with increasing redshift:
for an $S_{870}$ limit of $9-12.5$ mJy, the $\mstar$ limit ranges from $\sim 4-7 \times 10^{12} \msun$ at $z \sim 1$ to
$\sim 3-5 \times 10^{11} \msun$ at $z \sim 6$.
We discuss the few interferometrically detected SMGs that may be brighter than the proposed cutoff. Although such
objects are certainly interesting, inferences based on such objects may not apply to most SMGs.
\end{abstract}


\section{Introduction}

Explaining the observed number counts of high-redshift submillimetre (submm) galaxies (SMGs, which are traditionally defined as having
850-\micron~flux density $S_{850} \ga 3-5$ mJy; see \citealt{Blain:2002} and \citealt{Shapley:2011} for reviews)
has long been a challenging problem for hierarchical galaxy formation models. Thus, SMGs have attracted
much interest from theoreticians \citep[e.g.,][]{Baugh:2005,Fontanot:2007,Narayanan:2009,Narayanan:2010smg,Dave:2010,Hayward:2011smg_selection,
Hayward:2012smg_bimodality,Hayward:2013number_counts,Niemi:2012,Shimizu:2012,Somerville:2012}.
To explain the observed counts, some have resorted to significant modifications of their models, such as the use of a
top-heavy initial mass function \citep[e.g.,][]{Baugh:2005,Swinbank:2008,Dave:2010}.

It is becoming increasingly clear that single dish-detected SMGs are a heterogeneous population. Although some fraction of SMGs show strong
evidence for being merger-induced starbursts, a significant fraction seem to be early-stage mergers in which the components have projected
separation $\ga 10$ kpc (e.g., \citealt{Engel:2010}; for a thorough discussion of the observational evidence, see \citealt{Hayward:2012smg_bimodality}).
In \citet[][hereafter H13]{Hayward:2013number_counts}, we presented a model for the SMG population that includes the following physically distinct
subpopulations: 1. merger-induced starbursts; 2. early-stage mergers in which the galaxies are not yet strongly interacting but are blended into
a single submm source in single-dish surveys, which we term `galaxy-pair SMGs'; and 3. isolated discs. Furthermore, there is anecdotal evidence that
physically and spatially unassociated galaxies blended into a single submm source also contribute to the population \citep{Wang:2011}. However,
the contribution of the last subpopulation is currently very poorly constrained, and it is not included in the H13 model.

Except for H13, all published theoretical models do not treat the effects of blending. In the H13 model,
the galaxy-pair SMGs contribute significantly to the population at all fluxes, but merger-induced starbursts dominate at the
bright end. For $S_{870} \la 9$ mJy, the counts predicted by the H13 model are in good agreement with those of single-dish surveys, and the
Atacama Large Millimeter/submillimeter Array (ALMA) counts that we predicted have since been shown to agree very well with observations \citep{Karim:2012}.
However, for higher fluxes, the observed ALMA counts decrease significantly more steeply at the bright end than both the single-dish counts and the ALMA
counts predicted by H13 because the brightest ($S_{870} \ga 9$ mJy) single-dish-detected SMGs are almost all resolved into two or more sources.
This result is indeed very surprising. \citet{Karim:2012} suggest that their results imply an
upper limit on the star formation rate (SFR) of an SMG of $\sim 10^3 \msunperyr$ (for a \citet{Salpeter:1955} IMF; for a \citet{Kroupa:2001} IMF, the SFR
limit is a factor of $\sim 2$ lower). Based on a gas consumption time-scale argument that assumes that
SMGs are starbursts, they argue that this limit implies that the space density of galaxies with gas mass $\ga 5 \times 10^{10} \msun$
is $\la 10^{-5}$ Mpc$^{-3}$. In this Letter, we elaborate on this proposed limit by 1. using the results of dust radiative transfer calculations to
infer the SFR limit that corresponds to a range of flux limits and 2. presenting a redshift-dependent stellar mass limit that applies to all constituents of the SMG population,
not just starbursts. For simplicity, we treat the flux density limit suggested by observations as a strict upper limit, but in reality, it only implies an upper limit on the
number density of SMGs that are brighter than the limit (see \citealt{Karim:2012}).

The H13 predictions depend on the assumed stellar mass function (SMF), which is especially uncertain at high masses and high redshifts. Thus, it is potentially interesting
to remove the components of the H13 model that depend on the SMF and instead work `backwards', i.e., make inferences about SMG demographics from the observations.
In this Letter, we use the relationship among submm flux density, SFR, and dust mass predicted by our hydrodynamical simulations and dust radiative
transfer to infer an SFR limit from the cutoff flux density suggested by recent observations. Furthermore, because our model predicts submm flux density as
a function of stellar mass and redshift for quiescently star-forming galaxies (i.e., isolated discs and the individual components of `galaxy-pair' SMGs) -- and
at fixed stellar mass, starburst SMGs should have higher submm flux density than quiescently star-forming SMGs -- we can also predict an upper limit on the stellar mass of SMGs.

\section{The star formation rate and stellar mass limits implied by the potential flux limit}

\begin{figure}
\centering
\includegraphics[width=8cm]{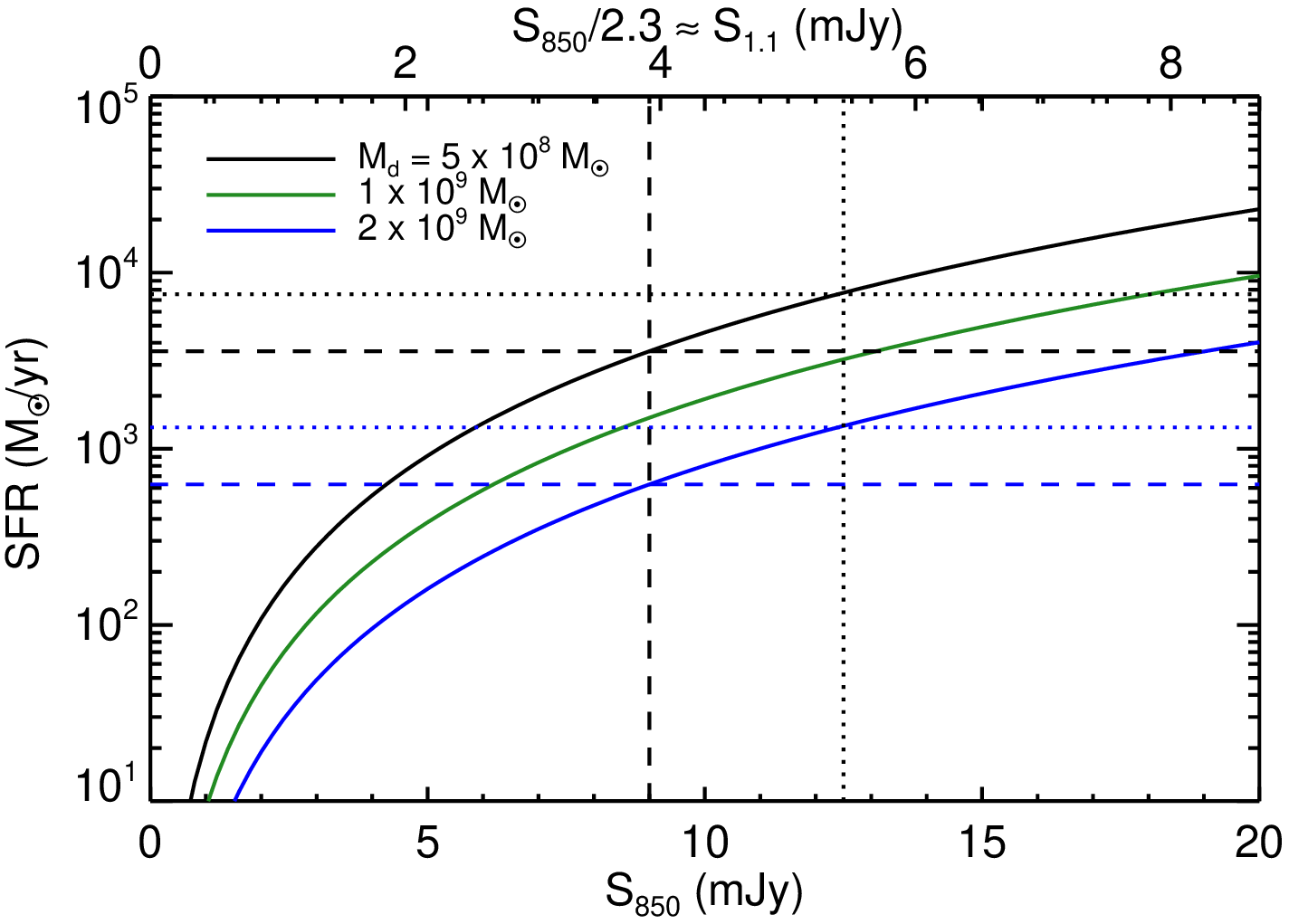}
\caption{Using the results of our hydrodynamical simulations and dust radiative transfer, the proposed limit on the submm flux density of SMGs \citep{Karim:2012,Smolcic:2012}
can be translated into a limit on the SFR.
This figure shows the relation between SFR ($\msunperyr$) and $S_{850}$ (mJy) calculated using Equation (\ref{eq:sfr_850}) for the $\mdust$ values given
in the legend. The vertical dashed (dotted) line indicates $S_{850} = 9~(12.5)$ mJy.
The horizontal blue (black) lines indicate the corresponding SFR limits for $\mdust = 2 \times 10^9~(5 \times 10^8) \msun$, which are
$\sim 630~(3600) \msunperyr$ for $S_{850} \la 9$ mJy and $\sim 1400~(7700) \msunperyr$ for $S_{850} \la 12.5$ mJy.}
\label{fig:sfr_vs_flux}
\end{figure}

Dust radiative transfer performed on hydrodynamical simulations of isolated disc galaxies and galaxy mergers suggests that the submm flux of a galaxy at $z \sim 1 - 6$
can be parameterized as a power law of the SFR and dust mass ($\mdust$) to within $\sim 0.15$ dex \citep[][H13]{Hayward:2011smg_selection}.
The physical reason underlying the parameterization is that the radiation and dust are in thermal equilibrium; the specific form can be motivated by a
toy model of single-temperature dust that is optically thin to its own emission \citep{Hayward:2011smg_selection}.
The parameterization is relatively insensitive to the dust geometry and the sub-resolution dust obscuration model, 
and it agrees well with the results of {\sc grasil} \citep{Silva:1998} calculations (A. Benson, private communication).
Consequently, one can infer the SFR of a galaxy from its submm flux, albeit with some uncertainty caused by the dependence on $\mdust$.
Solving for the SFR in equations (15) and (16) of H13 (which were derived from simulated SMGs with $S_{850} \la 15$ mJy, so extension to higher fluxes is
an extrapolation), we have
\begin{eqnarray}
\mathrm{SFR}_{850} &=& 9 \left(\frac{S_{850}}{\mathrm{mJy}}\right)^{2.33} \left(\frac{M_{\rm d}}{10^9 \msun}\right)^{-1.26} \msunperyr \label{eq:sfr_850}, \\
\mathrm{SFR}_{1.1} &=& 56 \left(\frac{S_{1.1}}{\mathrm{mJy}}\right)^{2.44} \left(\frac{M_{\rm d}}{10^9 \msun}\right)^{-1.37} \msunperyr, \label{eq:sfr_11}
\end{eqnarray}
where the subscripts of the SFR variables denote the flux from which the SFR is inferred and $S_{850}$ and $S_{1.1}$ are the
850-$\micron$ and 1.1-mm flux densities, respectively.\footnote{Note that these relations inherently assume a \citet{Kroupa:2001} IMF. Also,
for simplicity, we assume throughout this work that $ S_{860}$ and $S_{870}$ are identical to $S_{850}$ and use them interchangeably because the conversions
amongst them are insignificant for our purposes.}
The relationship between SFR and $S_{850}$ calculated using Equation (\ref{eq:sfr_850}) for various $\mdust$ values is shown in Fig. \ref{fig:sfr_vs_flux}.
Because $S_{850} \approx 2.3 S_{1.1}$ (H13), we simply show approximate $S_{1.1}$ values on the top axis of Fig. \ref{fig:sfr_vs_flux}.

Early interferometric follow-up of SMGs often focused on the brightest single dish-detected SMGs
\citep[e.g.,][]{Younger:2007high-z_SMGs,Younger:2008LH,Younger:2008,Younger:2009SMG_interf,
Smolcic:2011,Smolcic:2012CARMA}.
Only recently have interferometrically observed (sub)mm continuum maps of large (i.e., containing more than $\sim 10$), relatively unbiased samples of
less-luminous (i.e., $S_{850} \la 10$ mJy) SMGs become available. Most notably,
\citet{Karim:2012} mapped a sample of 122 SMGs from the LABOCA Extended \textit{Chandra} Deep Field South Submillimetre Survey (LESS; \citealt{Weiss:2009}; for the
description of the ALMA follow-up, see Hodge et al., submitted).
At all fluxes, a significant fraction (at least tens of per cent) of the SMGs were resolved into two or more components by ALMA.
As mentioned above, the brightest single-dish SMGs were all resolved into multiple sources; none of the resolved sources have $S_{870} \ga 9$ mJy.
The interferometrically imaged samples presented in \citet{Smolcic:2012} and \citet{Barger:2012} also tentatively support a steeper cutoff at the bright end than
expected based on the single-dish counts, although for these samples, the cutoff occurs at a higher flux density ($S_{870} \approx 12.5$ mJy), and there
are some sources that are potentially brighter than the cutoff (see Section \ref{S:outliers} for details).
The flux density at which the cutoff occurs may depend on the field because of cosmic variance (i.e., the highest overdensity sampled can vary from field to field)
or the different beam sizes of the interferometers used.

Because submm flux depends weakly on SFR \citep{Hayward:2011smg_selection}, the inferred SFR limit is sensitive to the maximum value of the
submm flux observed. The SFR limit is also sensitive to the value of $\mdust$.
The $\mdust$ values inferred from fitting the far-infrared spectral energy distributions of SMGs can be
in excess of $10^9 \msun$ but are often lower \citep[e.g.,][]{Magnelli:2012}. However, the accuracy to which $\mdust$ can be determined is still debated
\citep[e.g.,][]{Dale:2012}.
Thus, we will determine the SFR range that corresponds to $S_{870}$ cutoffs in the range $9 - 12.5$ mJy for a range of $\mdust$ values that are
characteristic of those inferred for SMGs \citep[e.g.,][]{Magnelli:2012}.
Fig. \ref{fig:sfr_vs_flux} indicates the $S_{870} = 9~(12.5)$ mJy limit
with a vertical dashed (dotted) line. The corresponding SFR limits for $\mdust = 2 \times 10^9~(5 \times 10^8) \msun$ are shown as blue (black)
horizontal lines. Assuming $\mdust = 2 \times 10^9~(5 \times 10^8) \msun$, $S_{870}$ limits in the range $9-12.5$ mJy correspond to maximum SFRs in the
range $\sim 630-1400~(3600-7700) \msunperyr$.
Note that the SFR limit for $\mdust = 2 \times 10^9 \msun$ is similar to that proposed by \citet{Karim:2012}.

\subsection{Quiescently star-forming SMGs} \label{S:quiescent_SF}

\begin{figure}
\centering
\includegraphics[width=8cm]{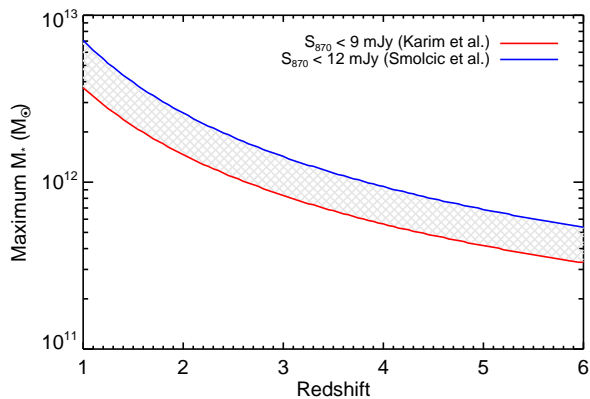}
\caption{The observed sharp cutoff in the SMG number counts suggests that there is a correspondingly sharp cutoff in the stellar mass function of star-forming galaxies at a given redshift.
The above plot shows the predicted $\mstar(z)$ limit that corresponds to an $S_{870}$ limit in the range $9-12.5$ mJy, where a higher $S_{870}$ limit
corresponds to a higher $\mstar$ limit. See the text for details regarding the calculation of the limit.}
\label{fig:max_mstar_vs_z}
\end{figure}

In the H13 model, both isolated disc galaxies (which dominate the faint, i.e., $S_{850} \la 2$ mJy, SMG population) and the galaxy-pair SMGs (i.e., early-stage mergers
in which the discs are not yet strongly interacting) form stars quiescently (as opposed to via the starburst mode).
The H13 relations for the redshift evolution of the gas content, size, and metallicity of galaxies can be used to predict
a quiescently star-forming galaxy's SFR and $\mdust$ as a function of its $\mstar$ and $z$. Because the (sub)mm flux of our simulated galaxies can be parameterized as a
function of SFR and $\mdust$ alone, the H13 model can be used to determine the (sub)mm flux as a function of $\mstar$ and $z$ (see equations 17 and 18
of H13). Thus for a given redshift, the SMG selection effectively selects the most massive quiescently star-forming galaxies on the main sequence
because the $S_{850}$--$\mstar$ relation predicted by our model for fixed redshift is monotonic (although in practice, scatter in the SFR--$\mstar$ relation will cause scatter in the
$S_{850}$--$\mstar$ relation; \citealt{Michalowski:2012}). Therefore, the observed flux density limits can be translated into upper limits on $\mstar$ as a function of $z$.

Fig. \ref{fig:max_mstar_vs_z} shows the inferred $\mstar$ limit versus $z$ for an $S_{870}$ limit in the range $9-12.5$ mJy.
This limit is calculated by combining the redshift-dependent SFR--$\mstar$ relation predicted by our model, which is tied to the dependence of galaxy gas fractions and sizes
on $\mstar$ and $z$, and the dependence of the dust mass on $\mstar$ and $z$, which depends on the gas fraction and metallicity evolution.
Then, the flux limit and the $\mstar$ and $\mdust$ functions are substituted into Eq. (\ref{eq:sfr_850}), and the resulting equation is solved for $\mstar$.
Because our model predicts an SFR--$\mstar$ relation that shifts upward as $z$ increases, the mass limit decreases with redshift.
Thus, it is natural to expect that the brightest quiescently star-forming SMGs are at higher redshifts than the median for the SMG population
if the high-mass end of the stellar mass function evolves less rapidly than the mass limit
plotted in Fig. \ref{fig:max_mstar_vs_z}.\footnote{This conclusion will not hold at redshifts beyond which the normalisation of the SFR-$\mstar$ relation no longer increases.
Observationally, the normalisation of the SFR--$\mstar$ relation seems to increase with redshift out to $z \sim 2$ \citep[e.g.,][]{Noeske:2007a,Daddi:2007}; whether the
trend continues to higher redshift is still an open question \citep[e.g.,][]{Gonzalez:2010,Rodighiero:2010,Karim:2011,McLure:2011,Whitaker:2012}.}
Indeed, we shall see below that most
of the very few sources with $S_{870} \ga 12.5$ mJy have redshifts significantly greater than the median redshift of the population.
The $\mstar$ limit that corresponds to an $S_{870}$ limit of $9-12.5$ mJy ranges from $\sim 4-7 \times 10^{12} \msun$ at $z \sim 1$ to $\sim 3-5
\times 10^{11} \msun$ at $z \sim 6$.

It is also interesting to consider the specific SFR, SSFR $=$ SFR$/\mstar$. In the H13 model, the $\mstar$ and $z$ dependence of the SSFR are determined by
the dependence of galaxy gas fractions and sizes on $\mstar$ and $z$. At fixed redshift, SSFR decreases with $\mstar$;
the scaling is similar to that of the star-forming galaxies in the \citet{Karim:2011} sample, SSFR $\propto \mstar^{-0.4}$. At fixed $\mstar$, SSFR increases with redshift. For example,
a galaxy with $\mstar = 10^{11} \msun$, approximately the minimum $\mstar$ necessary for a main sequence galaxy to contribute to the SMG population (see fig. 3 of H13),
has SSFR that ranges from 0.4~$\pergyr$ at $z \sim 1$ to 6.7~$\pergyr$ at $z \sim 6$. Consequently, we expect the higher-redshift main sequence SMGs to have higher
SSFRs. Furthermore, the H13 model suggests that main sequence SMGs should have SSFR $\la 7 \pergyr$. 

\subsection{Merger-induced starburst SMGs}

In the H13 model, merger-induced starbursts contribute significantly to the SMG population, and they dominate the bright (i.e., $S_{850} \ga 5$ mJy) end of the population.
Merger-induced starbursts can be outliers from the main sequence (i.e., have higher SFR for a fixed $\mstar$, but not all merger-induced starbursts are outliers from the main sequence;
\citealt{Hayward:2012smg_bimodality}).
Indeed, observations indicate that SMGs are a mix of main sequence members and outliers \citep{Michalowski:2012}. (Note that, as discussed below, some of the SMGs that are outliers
from the main sequence could also be gravitationally unstable discs.)
If the sources with flux density near the proposed cutoff flux density all lie significantly above the SFR--$\mstar$ relation,
then the upper limit on $\mstar$ presented in Section \ref{S:quiescent_SF} would overestimate the true $\mstar$ limit.

In the merger simulations of H13, which can account for the observed abundance of SMGs that are fainter than the possible flux limit,
the highest SFR attained is $\sim 5400 \msunperyr$. The peak SFR is typically limited by gas consumption rather than AGN
feedback\footnote{Note that the H13 simulations do not include starburst-driven winds. However, extremely high-resolution merger simulations \citep{Hopkins:2012mergers}
that include sophisticated prescriptions for feedback from supernovae, stellar winds, radiation pressure, and photoionisation
suggest that the peak SFR of the merger-induced starburst is relatively unaffected by starburst-driven winds \citep{Hopkins:2013merger_winds}.};
even without AGN feedback, an SFR in excess of $10^3 \msunperyr$ can only be sustained for a few $\times 10$ Myr unless fresh gas is supplied to the galaxy
at a comparable rate. Thus, the potential SFR limit discussed in this work may simply be a consequence of limited gas supply,
as argued by \citet{Karim:2012}, rather than evidence for strong feedback during the starburst.

Also, note that the starburst driven at coalescence in major mergers can elevate the SSFR beyond the aforementioned maximum for quiescently star-forming galaxies.
In simulations of mergers with baryonic mass ratio $\ga 1/3$, the SSFR can be boosted by $\sim 10-20 \times$ for a very short period ($\sim 30$ Myr at most)
and a factor of a few for as long as $\sim 100$ Myr.

\section{Sources that potentially exceed the flux density limit} \label{S:outliers}

\ctable[
	caption = {Sources that potentially exceed the flux density limit\label{tab:outliers}},
	center,
	notespar,
	doinside=\small
]{lcc}{
	\tnote[a]{For all but MM J120546-0741.5, the flux density is for a wavelength in the range $850-890~\micron$; for that source, the 1.2-mm flux density is given.}
	\tnote[b]{Reference(s): (1) \citealt{Iono:2006}, (2) \citealt{Younger:2009SMG_interf}, (3) \citealt{Wagg:2012}, (4) \citealt{Aravena:2010}, (5) \citealt{Younger:2007high-z_SMGs},
	(6) \citealt{Younger:2008}, (7) \citealt{Younger:2008LH}, (8) \citealt{Younger:2009SMG_interf}, (9) \citealt{Smolcic:2012CARMA}, (10) \citealt{Barger:2012}, (11) \citealt{Dannerbauer:2002}.}
}{
																				\FL
Identifier(s)					&	Flux density\tmark[a] (mJy)	&	Reference\tmark[b]		\ML
GN20 						& 	$22.9 \pm 2.8$		&	(1)			\NN
AzTEC8, Cosla-73, AzTEC/C2		&	$21.6 \pm 2.3$		&	(2)			\NN
BR1202-0725 North				&	$18.8 \pm 2.8$		&	(3)			\NN
BR1202-0725 South				&	$18.0 \pm 2.7$		&	(3)			\NN
MM1, Cosbo-1					&	$16.8 \pm 1.5$ 		&	(4)			\NN
AzTEC4, AzTEC/C4				&	$14.4 \pm 1.9$		&	(5)			\NN
AzTEC1, COSLA-60, AzTEC/C5	&	$13.8 \pm 2.3$		&	(5), (6)		\NN
LH 850.02					&	$13.4 \pm 2.4$		&	(7)			\NN
AzTEC2, COSLA-4, AzTEC/C3		&	$12.4 \pm 1.0$ 		&	(5)			\NN
AzTEC12, AzTEC/C18			&	$12.8 \pm 2.9$		&	(8)			\NN
AzTEC/C1, COSLA-89			&	$12.4 \pm 3.7$		&	(9)			\NN
AzTEC7						&	$12.0 \pm 1.5$		&	(5)			\NN
GN10, GOODS 850-5			&	$12.0 \pm 1.4$		&	(10)			\NN
MM J120546-0741.5			&	$6.5 \pm 0.8$		&	(11)			\LL
}

We have performed a literature search to identify SMGs for which interferometric continuum imaging indicates $S_{870} \ga 12.5$ mJy;
they are especially interesting targets for follow-up observations to determine whether they are indeed
brighter than the proposed flux limit and to discern the physical nature of these potentially extreme sources.
We are aware of seven sources at $z \ga 1$ (we are uninterested in low-$z$ interlopers)
that are at least 1$\sigma$ and eight additional sources that are possibly brighter than the $S_{870} \approx 12.5$ mJy cutoff.
The names of these sources, their flux densities, and references to the interferometric continuum observations of which we are aware that confirm the high fluxes
are presented in Table \ref{tab:outliers}.

Many of the sources that are potentially brighter than the proposed flux limit are from the AzTEC survey of a subfield of the
Cosmic Evolution Survey (COSMOS; \citealt{Scoville:2007}) field performed by \citet{Scott:2008}.
The number counts inferred by \citet{Aretxaga:2011} for a larger subfield that contains the \citet{Scott:2008} field are significantly higher
at the bright end than those derived from the similar-sized Submillimetre Common User Bolometer Array (SCUBA)
Half Degree Extragalactic Survey (SHADES) field \citep{Austermann:2010}. The brighter sources tend to coincide
with overdensities of galaxies at $z \la 1.1$, which may suggest that galaxy-galaxy lensing is responsible for the excess of bright
sources \citep{Austermann:2009,Aretxaga:2011}. Some of the other brightest sources may also be lensed.

However, it is unlikely that lensing is responsible for all of the sources that are potentially brighter than the proposed flux density limit, and it
is worthwhile to consider whether the sources that are brighter than the proposed limit are exceptional in some manner.
As we argued above, many of the brightest sources are at redshifts significantly greater than the median redshift of the SMG population
\citep[$\sim 2.6 - 3.5$; e.g.,][]{Yun:2012,Smolcic:2012,Weiss:2013}. For example, GN20 and GN10 are both located
in a $z = 4.04$ protocluster \citep{Daddi:2009smga,Daddi:2009smgb}, AzTEC1 is at $z = 4.6$ \citep{Smolcic:2011}, and
BR1202-0725 North and South are in a $z = 4.69$ overdensity. The scarcity
of bright sources at higher redshifts may be caused by a lack of sufficiently massive galaxies at those redshifts \citep{Hayward:2012thesis}
and/or insufficient dust content because of the limited time available to produce dust \citep[e.g.,][]{Michalowski:2010production}.
If indeed the bulk of the sources brighter than the proposed cutoff are at $z \ga 4$, this result may suggest that the normalisation of the
SFR--$\mstar$ relation continues to increase to at least that redshift; otherwise, there would be no clear reason
for the brightest sources to be preferentially at $z \ga 4$, and the evolution of both the stellar mass function and dust content of galaxies
should tend to make the brightest SMGs \textit{less} abundant at those redshifts.

As an example of what can potentially be learned from detailed follow-up of the sources that are potentially brighter than the proposed cutoff, we consider the case of GN20
\citep{Pope:2006}, which is an especially interesting and well-studied source. Recent high-resolution interferometric observations performed with the
Very Large Array (VLA) indicate that GN20 is a very extended ($\sim 14$ kpc in diameter) gas-rich disc that contains at least five molecular gas clumps, each of which
contain a few percent of the total molecular gas mass of the galaxy \citep{Hodge:2012}.
In our model, the redshift evolution of the SFR--$\mstar$ relation alone cannot account for the extremely high flux density of
GN20 (which does not seem to be lensed; \citealt{Carilli:2010}): Fig. \ref{fig:max_mstar_vs_z}
indicates that at $z \sim 4$, a galaxy with $\mstar \la 5 \times 10^{11} \msun$ would have $S_{870} \la 9$ mJy, and the inferred $\mstar$ value for
GN20 ($\mstar \sim 2 \times 10^{11} \msun$; \citealt{Daddi:2009smgb}) is less than this value.
Thus, GN20 seems to be an outlier from the main sequence \citep[see also][]{Daddi:2009smgb,Carilli:2010,Carilli:2011,Hodge:2012}:
its SSFR is $\sim 8.6 \pergyr$ \citep{Magdis:2011}, but in our model, a $z \sim 4$ main sequence galaxy with $\mstar = 2 \times 10^{11} \msun$ has SSFR $= 2.3 \pergyr$, $\sim 4 \times$
less than that of GN20. The simplest interpretation of GN20 is that it is an extremely massive, gas-rich, gravitationally unstable disc undergoing rapid fragmentation, but whether
such galaxies constitute a significant fraction of the SMG population is unclear.

\section{Summary and Discussion}

Recent interferometric imaging of single-dish-detected SMGs has demonstrated that most of the brightest ($S_{870} \ga 9 - 12.5$ mJy) sources are actually multiple distinct sources blended
into a single source. In this Letter, we have used the relationship among submm flux density, SFR, and dust mass predicted by our combination of hydrodynamical simulations and dust radiative transfer
and the redshift evolution of the SFR--$\mstar$ and $\mdust$--$\mstar$ relations predicted by the H13 model to predict limits on the SFRs and stellar masses of SMGs from the sharp cutoff in the
number counts suggested by recent observations. For $\mdust = 2 \times 10^9~(5 \times 10^8)
\msun$, an $S_{870}$ limit in the range $9-12.5$ mJy corresponds to an SFR limit in the range $\sim630-1400~(3600-7700) \msunperyr$.
The $\mstar$ limit ranges from $\sim 4-7 \times 10^{12} \msun$ at $z \sim 1$ to $\sim 3-5 \times 10^{11} \msun$ at $z \sim 6$.

Follow-up observations of the sources that are potentially brighter than the proposed flux limit are crucial for understanding whether these sources are exceptional in some
sense (besides their bright submm fluxes). Are they predominantly merger-induced starbursts, gravitationally unstable discs, or something else? Such
observations may yield new insight into star formation and feedback at high redshift.

However, one should be cautious when attempting to make inferences about the SMG population as a whole from observations of potentially extreme
objects such as GN20, which is one of the brightest SMGs known and, contrary to na\"ive expectations, is an extended very gas-rich disc galaxy rather
than a starburst \citep{Hodge:2012}.
Because the H13 model, which does not include violently unstable discs, can largely account for the observed SMG number counts,
it is not clear that the SMG population must include a significant number of such objects.
Even if some SMGs are massive, gas-rich, gravitationally unstable discs, this does not directly imply that they are fueled by `cold-mode accretion'
\citep{Birnboim:2003,Keres:2005}, as has been suggested by some authors \citep[e.g.,][]{Dave:2010,Carilli:2010,Hodge:2012}.
Rather, it simply requires that the discs are sufficiently gas-rich to be gravitationally unstable and is agnostic regarding the method of gas supply (see \citealt{Hopkins:2013cold_flow} for further discussion).

Finally, we note that we have not attempted to reconcile the discrepancy between the counts predicted by H13 and those of \citet{Karim:2012} in this work. One potential reason for the
discrepancy is that in H13, we did not include the effects of blending of physically and spatially unassociated galaxies. We also did not include mergers
of more than two galaxies. These potential contributions will be modeled in future work.

\acknowledgments

I thank Alex Karim, Jackie Hodge, Vernesa Smol\v{c}i\'c, and Volker Springel for insightful comments on the manuscript, Joel Primack for useful discussion, and
the anonymous referee for comments that led to a significantly improved manuscript. I am grateful to the Klaus Tschira Foundation for financial support.
\\

\footnotesize{
\bibliography{smg,std_citations}
}

\label{lastpage}

\end{document}